\documentclass[12pt]{article}

\input epsf
\usepackage{amssymb,wrapfig}
\usepackage[matrix,arrow,curve]{xy}
\usepackage{cite}
\usepackage{graphicx}
\usepackage{amsmath}

\makeatletter
\@addtoreset{equation}{section}
\makeatother

\def\nn {{\cal N}}

\def\cc {{\Bbb C}}
\def\pp {{\Bbb P}}
\def\zz {{\Bbb Z}}
\def\del {\partial}

\def\ka {K\"ahler}
\def\del {\partial}

\def\stt {{$\mathrm{SU(3)}\times\mathrm{SU(3)}$}}

\begin{document}

        \begin{titlepage}

        \begin{center}

        \vskip .3in \noindent

        {\Large \bf{The gauge dual of Romans mass}}

        \bigskip

        Davide Gaiotto$^1$ and Alessandro Tomasiello$^{2,3}$\\

        \bigskip

       {\small $^1$ School of Natural Sciences,
    Institute for Advanced Study, Princeton, NJ 08540, USA\\
    \vspace{.1cm}

        $^2$ Jefferson Physical Laboratory, Harvard University,
        Cambridge, MA 02138, USA
        \vspace{.1cm}

		$^3$ Universit\`a di Milano--Bicocca
        and INFN, sezione di Milano--Bicocca,
        I-20126 Milano, Italy}

        \vskip .5in
        {\bf Abstract }
        \vskip .1in

        \end{center}

        \noindent
We deform the recently proposed holographic duality between the ABJM ${\cal N}=6$ Chern-Simons-matter theory and type IIA string theory in AdS$_4\times \cc\pp^3$. We add a non--zero Romans mass $F_0$, whose dual we identify as the sum of the Chern--Simons levels for the two gauge groups. One can naturally identify four different theories, with different amounts
of supersymmetry and of flavor symmetry.

        \vfill
        \eject


        \end{titlepage}

\section{Introduction} 
\label{sec:intro}

The gauge/gravity duality proposed recently in
\cite{aharony-bergman-jafferis-maldacena} (ABJM) is the first example
in which the conformal field theory (CFT) is a three--dimensional theory with an explicit Lagrangian description.
The CFT$_3$
is a Chern--Simons theory coupled to matter fields. There are
two gauge groups, with Chern--Simons levels $k$ and $-k$.
The gravity duals are old
solutions \cite{nilsson-pope,sorokin-tkach-volkov-11-10} that involve
internal fluxes $F_2$ and $F_6$.

Using the Chern--Simons action to write down a Lagrangian for a
CFT is a very natural idea. Indeed, in three dimensions, the usual Yang--Mills action would involve a dimensionful coupling constant.
In contrast, the Chern--Simons action involves a dimensionless
parameters, which is even an integer for quantum--mechanical
reasons. Although the Chern--Simons theory is by itself
topological, and it hence describes only a finite number of degrees
of freedom, it can be coupled to scalars and fermions to achieve
scale invariance.

It was initially expected that such Chern--Simons--matter actions
would be dual to AdS$_4$ solutions in string theory involving
Romans mass $F_0$ \cite{schwarz-chernsimons}. This is
because of the coupling $\int_{\rm D2} F_0 CS(a)$, where $a$
is the world--volume gauge field on the D2.

The duals found in \cite{aharony-bergman-jafferis-maldacena}
confounded this expectation, in that no $F_0$ is present in the
solution, even though the theory contains a Chern--Simons action.
The reason for this is, roughly, that the internal flux $F_2$
also induces a Chern--Simons coupling on ``fractional D2's''
-- D4's wrapping a vanishing two--cycle.

This does not, however, invalidate the original reason to expect
that $F_0$ would induce a Chern--Simons action. In fact, supersymmetric
solutions with non--zero Romans mass were found recently in \cite{t-cp3}, and with an internal space whose topology is $\cc\pp^3$,
the same as in the ${\cal N}=6$ solutions \cite{nilsson-pope,
sorokin-tkach-volkov-11-10}. Since these solutions have only ${\cal N}=1$ supersymmetry, one might be skeptical of any proposed gauge duals to them. However, the solutions in \cite{t-cp3} happen to have a parameter space that, although discretized by flux quantization, gets arbitrarily
close to the ${\cal N}=6$ solutions of \cite{nilsson-pope,
sorokin-tkach-volkov-11-10}.

Exploiting this fact, in section \ref{sub:ft1} we will be able to
find Chern--Simons--matter theories which are, in a sense, small
deformations of the ABJM theory, and that should be dual to the
solutions in \cite{t-cp3}. These duals vindicate the expectation
mentioned above, that the Romans mass should contribute to the
Chern--Simons levels. In fact, both $F_0$ and $F_2$ contribute to
the Chern--Simons actions for the two gauge groups: morally, one
gauge group has level $k$, and the other  has level $F_0-k$, where
$k=\int_{\cc\pp^1}F_2$. See also (\ref{eq:levelsRomans}) and
(\ref{eq:match}) below.

In fact, in section \ref{sec:romans} we will show that this
is a general feature, dictated by considerations involving
brane probes and simple but robust topological arguments.

Also, we will see in section \ref{sec:multi} that similar ideas
to the ones that lead to the definition of the ${\cal N}=1$
field theory in
\ref{sub:ft1} also suggest how to define theories with ${\cal N}=0, 2$ and $3$.
At the same time, the SO(6)
R--symmetry of the ${\cal N}=6$ theory gets broken in various ways,
summarized in table \ref{tab:theories}.

This means that the situation is now reversed: it is now
field theory that suggests where to look for gravity solutions.
We discuss these duals in section \ref{sec:gravity}.
The ${\cal N}=1$ solutions, as we mentioned, were found in \cite{t-cp3}, and we review them in section \ref{sub:gr1};
but the solutions with ${\cal N}=0,2$ and 3 seem to be new.
Perhaps surprisingly, the ones which are easiest to find are
the solutions with ${\cal N}=0$, as we will see in section
\ref{sub:gr0}. This is because they have the
highest amount of isometries (as one can see from table \ref{tab:theories}). We have not found yet\footnote{There exist
${\cal N}=2$ solutions with an internal space whose topology
is $\cc\pp^3$, but they are vacua of the same effective theory that also describes the ${\cal N}=6$ solutions; hence, they cannot involve
$F_0\neq 0$. Three--dimensional analogues of the four--dimensional
``beta--deformed'' theories \cite{lunin-maldacena} might also be
considered, as done for example in \cite{imeroni}, but
such theories should be continuous deformations of the ${\cal N}=6$
theory, and hence should have the same Chern--Simons levels.}
 the solutions with
${\cal N}=2$ and 3, but we know a few features that they should
have, as we discuss in sections \ref{sub:ft2} and \ref{sub:ft3}.
Work is in progress to find them explicitly.


\section{Romans mass as sum of Chern--Simons levels} 
\label{sec:romans}

The Chern--Simons--matter theory found in \cite{aharony-bergman-jafferis-maldacena}, to be reviewed
more thoroughly in section \ref{sec:multi}, has two gauge groups,
with CS levels $k_1=-k_2\equiv-k$. It is the dual to a certain IIA
solution on AdS$_4\times \cc\pp^3$; we will refer to it as
the ``${\cal N}=6$ solution'' from now on.
Suppose one has a solution obtained by a small perturbation
of this ${\cal N}=6$ solution and involving a non--zero Romans
mass $F_0$. In this section, we will argue rather generally
that the dual theory to such a solution is a perturbation of
the ABJM theory with levels
\begin{equation}\label{eq:levelsRomans}
    k_1 + k_2 = F_0 \ .
\end{equation}

The argument relies on the field--operator correspondence in
AdS/CFT. In particular, we want to consider the field theory
duals to particles in AdS obtained by wrapping D--branes on
subspaces of the internal $\cc\pp^3$. Although the argument
ultimately only needs D0 branes, we will start by reviewing some
background material.

First of all, let us recall the better--known case of
AdS$_5 \times S^5$ \cite{witten-baryons}.
In that case, the $N$ quanta of
$F_5$ on the sphere induces a tadpole $\int a F_5 = N \int a$
for the U$(1)$ gauge field on
a D5 brane wrapping the $S^5$. This tadpole requires $N$
fundamental strings to end on the $D5$ brane, which is then
identified with a ``baryonic operator'', i.~e.~$N$ fundamental Wilson
lines ending on an $\epsilon$ tensor.

In the ABJM case, there is a richer story. The
$\nn=6$ solution on AdS$_4\times \cc\pp^3$
\cite{nilsson-pope,sorokin-tkach-volkov-11-10}
is characterized by two flux integers, not one: the
integrals of $F_2$ and $F_6$, $k\equiv\int_{\cc\pp^1} F_2$ and
$N\equiv \int_{\cc\pp^3}F_6$. In our notation (to be
reviewed more extensively in section \ref{sub:gr1}), these fluxes
are related to the overall radius $R$ in string units and to the
string coupling $g_s$ by
\begin{equation}
    k= \frac{\pi R}{g_s} \ ,\qquad N= \frac{\pi^3 R^5}{2 g_s}\ .
\end{equation}

In this solution, there are now two types of branes with tadpoles.
A D6 brane wrapping the whole $\cc\pp^3$ would have a
tadpole of $N$ units, require $N$ strings to end on it, and
correspond to $N$ Wilson lines ending on an $\epsilon$ tensor.
This baryonic vertex is quite similar to the one in AdS$_5 \times
S^5$.

Another brane with a tadpole is a D2 brane wrapping a $\cc\pp^1$;
such a brane has $k$ units of
U$(1)$ charge on its worldvolume.
In this case, the $k$ Wilson lines cannot
end on an $\epsilon$ tensor. However, in a Chern--Simons theory
at level $k$, a Wilson line in the representation Sym$_k$
(the one obtained by symmetrizing $k$ fundamentals), extended say
from infinity to a point $p$, is
equivalent \cite{moore-seiberg} to a monopole ('t Hooft) operator creating
one unit of flux around the point $p$. In other words, only
the endpoint of the Wilson line is physical. Hence, we can have
$k$ Wilson line end in a point, and this is the field theory dual
to $k$ fundamental strings ending on a D2 brane.

On the other hand, there are also branes that do not have any
tadpoles on their worldvolumes. For example, this is the case for
D4 branes. They have been
mapped to the so--called dibaryons, made of $N$ bifundamental fields $X_A$ attached to two $\epsilon$ tensors. Notice that the
mass of a D4 goes indeed like $m_{D4} R\sim R^5/g_s\sim N$.

Finally, D0 branes also have no tadpole, and
correspond to di--monopole operators, with
charge $(1,1)$ under the two gauge groups. Once again following
\cite{moore-seiberg}, because of the Chern--Simons couplings
of the two gauge groups,
these operators are equivalent to Wilson lines in the
representation $({\rm Sym}_k, \overline{{\rm Sym}_k})$.
In other words, these operators carry
both $k$ fundamental indices of one group, and $k$ anti--fundamentals
of the other group. These indices can be saturated by $k$
bifundamental $X_A$ fields. The mass of a D0 goes like
$m_{D0}R \sim R/g_s \sim k$, which again makes sense.

This completes our list of duals to D--branes for the original ABJM
duality. Let us now see how we can modify it. A simple modification
of the background consists in adding closed $B$--field. In general,
the most sensible definition of flux is the so--called ``Page
charge'', the integral on a $p$--cycle of $\tilde F_p \equiv (e^B
F)_p$. For example, one can generate in this situation a $\tilde F_4
= B F_2$ that can have a non--zero flux $n_4$.  In such a
background, the D4 brane has now $n$ units of U$(1)$ tadpole
(because $\tilde F_4$ is the flux coupling to the worldvolume gauge
field). We saw before that the dual to the D4 when $n_4=0$ was the
dibaryon operator. Now, for $n_4 \neq 0$, the tadpole is telling us
that this di--baryon operator has now $n$ ``dangling'' fundamental
indices on one of the two sides that have been contracted with
$\epsilon$ tensors. This suggests \cite{aharony-bergman-jafferis}
that the theory has now changed: the two ranks should now
differ by $n_4$.

One can now try to apply the same reasoning to the D0. We just saw
an easy way to modify the ABJM background by adding $\tilde F_4$ flux.
\footnote{Incidentally, one cannot add an
untilded $F_4$ without also adding $F_0$.} Let us now suppose we have
a way of deforming the background and of introducing a non--zero
Romans mass $F_0$; we have not shown how to do this at this point,
but we will see it later. Now a D0 brane will develop a tadpole,
because of the coupling $\int a F_0$, where $a$ is the worldvolume
U$(1)$ field. This means that one needs $F_0$ fundamental strings
ending on the D0. This can be explained with the two
Chern--Simons levels being no longer equal. Indeed, in such a
theory a di--monopole operator of charge $(1,1)$ has
$k_1$ indices in the fundamental of the first gauge group,
and $-k_2$ indices in the anti--fundamental of the second gauge
group. Even if one dresses this operator up with $k_2$ bifundamental
fields, one is left with $k_1 + k_2$ ``dangling'' fundamental
indices. These indices are in correspondence with the $F_0$
fundamental strings above. This leads us to conclude
(\ref{eq:levelsRomans}).
 
We can strengthen this conclusion by analyzing the field theory dual
of configurations where a single D2 brane, point-like in $\cc
\pp^3$, acts as a domain wall in $AdS_4$. These configurations,
which we will also consider in sections \ref{sub:ft2} and \ref{sub:ft3}, correspond in
the ABJM theory to vacua where the bifundamental fields $X^I$
in a single $U(1) \times U(1)$ block are given large expectation
values. The relevant terms in the Lagrangian are
\begin{equation}
k_1 CS(a_1) + k_2 CS(a_2) + |X|^2 (a_1 - a_2)^2
\end{equation}
The bifundamental fields are charged under the difference of
the two $U(1)$ gauge fields $a_- = a_1 - a_2$, which becomes very massive. At low
energy, the two gauge fields are forced to be equal. 

If the Chern--Simons levels are equal and opposite, the Chern--Simons
actions cancel each other. The dynamics of the surviving gauge field $a_+= a_1 + a_2$ 
 is controlled by a Yang--Mills term generated by integrating away the difference between the
gauge fields \cite{mukhi-papageorgakis}. Indeed the Lagrangian can be rewritten schematically  as
\begin{equation}
k a_- \wedge da_+ + |X|^2 a_-^2
\end{equation}
and $a_-$ integrated out explicitly to give 
\begin{equation}
\frac{k^2}{|X|^2} (da_+)^2
\end{equation}
$a_+$ matches well the gauge field living on the D2 brane domain
wall.

If the Chern--Simons levels are changed to more generic values, the
light gauge field $a=a_1=a_2$ will have a residual $k_1 + k_2$ Chern--Simons
coupling. This matches the above mentioned coupling $\int_{D2} F_0
CS(a)$ on a D2 brane in the presence of Roman mass if $F_0 = k_1 +
k_2$.

Having reached this general conclusion, we now have to ask
ourselves whether one can define CFTs by deforming the ABJM theory
by letting the ranks not sum to zero, and -- dually -- whether
there are any supergravity solutions that deform the ${\cal N}=6$
solution by a small non--zero Romans mass $F_0$. We will
consider these two questions in sections \ref{sec:multi}
and \ref{sec:gravity}.


\section{A hierarchy of theories with unequal levels} 
\label{sec:multi}

In this section, we start from the ABJM theory and try to
let the levels of the two gauge groups not sum to zero any more:
$k_1 \neq - k_2$.

This is a bit ambiguous, however. By supersymmetry, the level
appears in several terms in the ABJM action, such as the
bosonic potential and the fermion--boson couplings. How many
of those coefficients we change is up to us.
As we will explain in this section, we can choose to preserve different amounts of supersymmetry and different global symmetries.
The theories we will define
are summarized in table \ref{tab:theories}.

\begin{table}[h]

\begin{center}
    \renewcommand{\arraystretch}{1.5}
    \begin{tabular}{c||c|c|}\cline{2-3}
    &supersymmetry & global symmetry \\\hline\hline
    \multicolumn{1}{|c||}{Sec.~\ref{sub:ft0}}& ${\cal N}=0$ &SO(6)\\\hline
    \multicolumn{1}{|c||}{Sec.~\ref{sub:ft1}}& ${\cal N}=1$ &SO(5)\\\hline
    \multicolumn{1}{|c||}{Sec.~\ref{sub:ft2}}& ${\cal N}=2$ &SO(2)$_{\rm R}\times$ SO(4)\\\hline
    \multicolumn{1}{|c||}{Sec.~\ref{sub:ft3}}& ${\cal N}=3$ &SO(3)$_{\rm R}\times$ SO(3)\\\hline
    \end{tabular}
\end{center}
\vskip 0.3cm
\label{tab:theories}
\small{{\bf Table 1:} The various theories with $k_1 \neq -k_2$
that we will define in this section. The subscript $_{\rm R}$ denotes R--symmetry.
}
\end{table}

We will argue that there is a CFT in each of these classes
of theories. In other words, for $k_1 \neq - k_2$, the ABJM
fixed point of the RG flow splits in four different fixed points,
each with a different amount of supersymmetry and global symmetries.

As we will see, there is more control on the field theory side as supersymmetry increases; on the gravity side,
to be discussed in section \ref{sec:gravity}, there is more
control as the amount of bosonic symmetry increases.

\subsection{${\cal N}=0$} 
\label{sub:ft0}

One can start by simply taking the ABJM written down in components
(as can be found for example in \cite[Sec.~4]{benna-klebanov-klose-smedback}), and
changing the Chern--Simons
levels, so that now $k_1 \neq -k_2$ -- and nothing else.

Once we do this, the theory is not conformal any more.
However, the RG flow will respect the SO(6) symmetry of the action.
In other words, the flow will happen in the finite--dimensional
space of theories QFT$_{{\cal N}=0}(k_1,k_2)$, defined as
the space of Chern--Simons--matter theories with the same
field content as the ABJM, with Chern--Simons levels $k_1$ and $k_2$
for the two gauge groups, and with SO(6) symmetry. For example,
the most general single--trace potential will have the form
\begin{equation}\label{eq:bospot}
    V=c_{IJK}^{LMN}
    {\rm Tr}[ X^I X_L^\dagger X^J X_M^\dagger X^K X_N^\dagger] \ ,
\end{equation}
where we have collected together all the fields in $X_I = (A_1, A_2,
B_1^\dagger, B_2^\dagger)$. The coefficients $c_{IJK}^{LMN}$ are
some of the coordinates on the space of theories QFT$_{{\cal
N}=0}(k_1,k_2)$. More generally, let us denote by $c_O$ the
coefficient multiplying the operator $O$; besides the terms in the
bosonic potential (\ref{eq:bospot}), there are quartic
fermion--boson couplings, and possible double--trace terms and
 triple--trace terms that we will discuss in due course. We
will denote by $c^0_O$ the value of the coefficient $c_O$ in the
ABJM theory.

We now want to argue that there will still be a CFT in this space
of theories, if $k_1 + k_2$ is small in a sense that we will specify.

Quantities as the $\beta$ functions of the $c_O$ will be functions
of the two t'Hooft couplings $\lambda_i = \frac{N}{k_i}$. For Chern
Simons theories, these t'Hooft couplings are not subject to RG flow.
We are interested in a regime where $\lambda_+ = \lambda_1 +
\lambda_2$ is kept small, while $\lambda_- = \lambda_1 - \lambda_2$
may be large.

If we change the levels and no other coefficient in the action, so
that $c_O = c^0_O$, we expect that the beta functions will not
change much with respect to the ABJM case, where they are $0$.
\begin{equation}
    \beta_O = (\lambda_1 + \lambda_2) \delta \beta_O \  .
\end{equation}

Essentially, we are assuming that the $\beta$ functions should be
analytic in $\lambda_+$ around $\lambda_+=0$, even for large
$\lambda_-$. In general $\delta \beta$ will not be zero, and the
theory will not be conformal. However, we can respond by changing
also the coefficient $c_O$ a bit. The variation of the beta function
is related to the anomalous dimension of the operator (for example,
a marginal operator can be added infinitesimally to a CFT without
changing the beta functions at the first order). Hence we can write:
\begin{equation}
    \beta_O = (\lambda_1 + \lambda_2) \delta \beta_O + \delta c_O \gamma_O\
\end{equation}
where $\delta c_O = c_O - c_O^0$
Now, if $\gamma_O \neq 0$, we can just take
\begin{equation}\label{eq:varbeta}
    \delta c_O = - (\lambda_1 + \lambda_2)\frac{\delta \beta_O}{\gamma_O}
\end{equation}
and obtain $\beta_O=0$ again.

Hence, in the space QFT$_{{\cal N}=0}(k_1,k_2)$, there will be a
conformal theory that we will denote CFT$_{{\cal N}=0}(k_1,k_2)$,
with Chern--Simons action with levels $k_1 \neq -k_2$, and the rest
of the action of the form $\sum_O (c_O + \delta c_O) O$.

There are a few things to remark about this logic. First of all,
looking at (\ref{eq:varbeta}), we can now say in what sense
$\lambda_1 + \lambda_2$ needs to be small: we have to assume
\begin{equation}\label{eq:small}
    (\lambda_1 + \lambda_2) \delta \beta_O \ll \gamma_O \
\end{equation}
for all $O$ that have SO(6) symmetry. At weak coupling, this means
that $k_1 + k_2 \ll k_1, k_2$. At strong coupling, we expect that
$(\lambda_1 + \lambda_2) = \frac{k_1 + k_2}{k_1 - k_2} \lambda_- \ll
1$ to be sufficient, though it is possible that the detailed
$\lambda_-$ dependence of $\delta \beta_O, \gamma_O$ will require a
more stringent $\frac{k_1 + k_2}{k_1 - k_2} \lambda_-^C \ll 1$ for
some constant $C>1$.

Second, our reasoning works only if
\begin{equation}\label{eq:nonmarg}
    \gamma_O \neq 0  \ ,
\end{equation}
again for all $O$ that are SO(6) singlets. If the operator $O$ is
marginal in the ABJM theory, $\gamma_O$ will also be proportional to
$\lambda_+$, and any attempt to set the beta function to zero will
require a large $\delta c_O$, outside the range of our perturbative
approximation.

It is easy to write down an operator of classical dimension 3, but
most of these will acquire an anomalous dimension in the ABJM
theory. The ones that remain marginal are the ones that are
protected by supersymmetry. To find these, one can determine the
chiral primaries for the ABJM theory. The single trace ones are of
the form ${\rm Tr}(X^{(I_1}X_{(J_1}^\dagger X^{I_2}X_{J_2}^\dagger
\ldots X^{I_n)}X_{J_n)}^\dagger)$,
 where the indices
up and down are separately symmetrized. This is the representation
$(n,0,n)$ of SU(4). (This result is similar to the computation of
chiral primaries for the super Yang--Mills theory in four
dimensions, that are in the $(0,n,0)$ of SU(4).) We can consider
arbitrary descendants of such chiral primaries. The $SO(6)$ content
of all such protected descendants can be found for example in
\cite[Table 1]{nilsson-pope} (the table lists the protected spectrum
of supergravity on $AdS_4 \times \cc\pp^3$, which coincides with the
spectrum of protected operators in the ABJM theory).

In any case, none of these protected operators of dimension $3$ are
SO(6) singlets. Hence our assumption (\ref{eq:nonmarg}) was
justified.

Now that we know that there is a CFT in the space of theories
QFT$_{{\cal N}=0}(k_1,k_2)$, we can also ask whether it is
attractive. For this, we need to know whether there are any relevant
operators among the $O$ that we allowed in the action (the ones
which are SO(6) singlets). This can be answered by looking for
relevant operators in the ABJM theory, since our new CFT is a small
deformation of it. This can in turn be analyzed using AdS/CFT:
relevant operators should correspond to tachyonic operators. We can
now check what these tachyonic operators are from the same table
\cite[Table 1]{nilsson-pope}. In that table, we can see that the
only scalars with negative mass squared\footnote{The mass conventions used
in \cite{nilsson-pope} are given in units of $1/R^2$, and after
subtracting a term to make the massless case conformally invariant
\cite[3.2.22]{duff-nilsson-pope}: the relation with the more usual
definition is $M^2_{\rm here}= \frac1{R^2}(\frac14({\rm
mass}_{NP})^2-2)$. The formula for conformal dimensions is then the
usual $\Delta(\Delta-3)= M^2_{\rm here}$.\label{foot:dnp}} are
contained in the rows denoted by $0^{+(1)}$ and $0^{-(1)}$,
respectively for $p=0,1$ (dual to operators of dimension
$\Delta=1,2$) and $p=0$ (dual to operators of dimension $\Delta=2$).
The cases $p=2$ in $0^{+(1)}$ and $p=1$ in $0^{-(1)}$ are massless
fields, dual to marginal operators. None of these operators is a
singlet. Hence, there are no relevant operators which are singlets
of SO(6) and neutral under the U(1) of M--theory.

To summarize, we have concluded that there exists a conformal field
theory CFT$_{{\cal N}=0}(k_1,k_2)$, with ${\cal N}=0$ and SO(6)
symmetry, and that this CFT is attractive in the space
QFT$_{{\cal N}=0}(k_1,k_2)$
of field theories with this symmetry and these Chern--Simons levels.

The analysis so far ignored multi-trace interactions. We will
discuss their role in section \ref{sub:Multitrace}.


\subsection{${\cal N}=1$} 
\label{sub:ft1} Another possibility is trying to impose ${\cal N}=1$
supersymmetry. The ABJM Lagrangian can be written in terms of ${\cal
N}=1$ superfields as
\begin{equation}\label{eq:1abjm}
    \begin{array}{l}\vspace{.3cm}
    S_{\rm ABJM}= \frac k{4\pi} (S_{\rm CS,\,{\cal N}=1}({\cal A}_1)-S_{\rm CS,\,{\cal N}=1}({\cal A}_2)) + \\
    \int d^2\theta\,{\rm Tr}\Big(D_a X_I^\dagger D^a X^I +
    \frac{2\pi}k (X_I^\dagger X^I X_J^\dagger X_J
    - X_I^\dagger X^J X_J^\dagger X^I -2
    \omega^{IK}\omega_{JL}X_I^\dagger X^J X_K^\dagger X_L )
    \Big)\ .
    \end{array}
\end{equation}
Here $X_I$ are now ${\cal N}=1$ superfields; we keep the same
notation as for the bosonic fields in the previous subsection.
The covariant derivatives are defined using the vector superfields
${\cal A}_i$.
The last three terms are the ${\cal N}=1$ superpotential (which
is a real function of the fields). It contains the symplectic
matrix $\omega$, defined as being
${{\epsilon\  0}\choose {0 \ \epsilon}}$, with
$\epsilon$ the antisymmetric 2$\times$2 matrix.
Notice that (\ref{eq:1abjm})
 makes explicit the Sp(2) invariance of the theory.

This time we define QFT$_{{\cal N}=1}(k_1,k_2)$ by
letting the two levels in (\ref{eq:1abjm}) be unequal,
and by letting the coefficients of the operators vary:
\begin{equation}
    \begin{array}{l}\vspace{.3cm}
    S_{{\rm QFT}_{{\cal N}=1}(k_1,k_2)}= \frac{k_1}{4\pi} S_{\rm CS,\,{\cal N}=1}({\cal A}_1)
    + \frac{k_2}{4\pi} S_{\rm CS,\,{\cal N}=1}({\cal A}_2) + \\
    \int d^2\theta \,{\rm Tr}\Big(D_a X_I^\dagger D^a X_I +
    (c_1 X_I^\dagger X^I X_J^\dagger X_J
    + c_2 X_I^\dagger X^J X_J^\dagger X^I + c_3
    \omega^{IK}\omega_{JL}X_I^\dagger X^J X_K^\dagger X_L )
    \Big)\ .
    \end{array}
\end{equation}
This is the most general ${\cal N}=1$ action with levels
$k_1 \neq -k_2$ and SO(5)$=$Sp(2) symmetry.

To establish whether there is a CFT in this space of theories, we
can argue just like in the previous section. Once again we have to
worry about the possible presence of marginal protected operators.
Now we need to consider operators of dimension $2$ integrated over
the ${\cal N}=1$ superspace. We can consult the usual table
\cite[Table 1]{nilsson-pope} and decompose
the SO(6) representations under
Sp(2). None of the representations of dimension $2$ or smaller
contains an Sp(2) singlet.

Hence, we can conclude that, in the limit (\ref{eq:small}),
there is a CFT with Chern--Simons levels $k_1 \neq -k_2$, with
${\cal N}=1$ supersymmetry and SO(5) global symmetry; and that
it is an attractive fixed point in the space of theories
with the same symmetries.


\subsection{${\cal N}=2$} 
\label{sub:ft2}

The ABJM theory can be written using
${\cal N}=2$ ``chiral'' superfields $A_i, B_i$ and
vector superfields $V_1, V_2$. The action then reads
\begin{equation}\label{eq:2abjm}
    \begin{array}{l}\vspace{.3cm}
        S_{\rm ABJM}=
        \frac k{4 \pi}(S_{{\rm CS},\,{\cal N}=2}(V_1) -
        S_{{\rm CS},\,{\cal N}=2}(V_2) ) +
    \int d^4 \theta {\rm Tr} (  e^{-V_1} A_i^\dagger e^{V_2} A_i +
        e^{-V_1} B_i e^{V_2} B_i^\dagger) \\
    \hspace{1cm}+ \left(\frac {2\pi} k\int d^2\theta\, \epsilon^{ij}\epsilon^{kl}A_i B_k A_j B_l  + {\rm c.\,c.}\right)
    \end{array}
\end{equation}
where $d^2 \theta \equiv d^2(\theta^1 + i \theta^2)$, and the
last line is the ${\cal N}=2$ superpotential (which is a holomorphic
function of the ${\cal N}=2$ chiral superfields).

Once again we modify the levels, but this time in an ${\cal N}=2$
sense. We also want to keep the SO(4) invariance which is manifest
in (\ref{eq:2abjm}). This leads us to
\begin{equation}\label{eq:cft2}
    \begin{array}{l}\vspace{.3cm}
        S_{{\rm QFT}_{{\cal N}=2}(k_1,k_2)}=
        \frac {k_1}{4 \pi} S_{{\rm CS},\,{\cal N}=2}(V_1) +
        \frac {k_2}{4 \pi} S_{{\rm CS},\,{\cal N}=2}(V_2) ) +
    \int d^4 \theta {\rm Tr} (  e^{-V_1} A_i^\dagger e^{V_2} A_i +
        e^{-V_1} B_i e^{V_2} B_i^\dagger) \\
    \hspace{1cm}+ \left(c
    \int d^2\theta\, \epsilon^{ij}\epsilon^{kl}A_i B_k A_j B_l  + {\rm c.\,c.}\right)\ .
    \end{array}
\end{equation}
This time, there is no other superpotential we can write, and all we
can allow to vary is the coefficient $c$ in the second line of
(\ref{eq:cft2}).  At weak coupling, we can actually use the logic in
\cite{gaiotto-yin}, that goes as follows. The beta function for the
coefficient $c$ should be positive for $c \gg 0$, since in that
limit the Chern--Simons action is small with respect to the action
of the scalars, and theory is essentially a Wess--Zumino model. At
$c \ll 0$, on the other hand, one can compute the beta function for
$c$ by computing the R--charges of the fields in the theory without
a superpotential. The result is that the beta function is negative.
Hence, in the IR, the coupling $c$ grows when it is small, and
decreases when it is large. This implies that there is a critical
value in between.

At strong coupling, we could once again argue that for sufficiently
small $\lambda_+$ there will be a fixed point for a value of $c$
close to the ${\cal N}=6$ value. As in subsections \ref{sub:ft0} and
\ref{sub:ft1} we can also check for the existence of marginal or
relevant $SO(4)$ invariant chiral primaries in the spectrum. Their
absence guarantees that the fixed point is attractive.

\subsection{${\cal N}=3$} 
\label{sub:ft3}

This is case is different from all the others, in that
one can actually write down the explicit Lagrangian: it is the
standard ${\cal N}=3$ theory with the assigned Chern--Simons levels
and field content. Using again ${\cal N}=2$ superfields\footnote{One could alternatively use ${\cal N}=3$ superfields; the ABJM action was written in this form in \cite{buchbinder-ivanov-lechtenfeld-pletnev-samsonov-zupnik}.}:
\begin{equation}\label{eq:cft3}
    \begin{array}{l}\vspace{.3cm}
        S_{{\rm CFT}_{{\cal N}=3}(k_1,k_2)}=
        \frac {k_1}{4 \pi} S_{{\rm CS},\,{\cal N}=2}(V_1) +
        \frac {k_2}{4 \pi} S_{{\rm CS},\,{\cal N}=2}(V_2) ) +
    \int d^4 \theta {\rm Tr} (e^{-V_1} A_i^\dagger e^{V_2} A_i +
    e^{-V_1} B_i e^{V_2} B_i^\dagger) \\
    \hspace{1cm}+ {2\pi}
    \int d^2\theta \Big(\frac 1{k_1}{\rm Tr}(B_i A_i)^2 +
    \frac 1{k_2}{\rm Tr}(A_i B_i)^2 \Big)\, \ .
    \end{array}
\end{equation}
In this case, the global symmetry is made up of the SO(3)$_R$,
and of the SO(3) that rotates the $A_i$ and $B_i$ simultaneously.


\subsection{Relations between different fixed points} 
\label{sub:ft4} Besides the four special fixed points which we
analyzed in the previous four subsections, there are other possibly
interesting theories which deform the ${\cal N}=6$ theory in less
symmetric directions.

As an example, consider the most general ${\cal N}=2$ single trace
deformation with a Pilch-Warner type $SU(2)$ invariant
superpotential \cite{Corrado:2002wx,
corrado-halmagyi,benvenuti-hanany,halmagyi-pilch-romelsberger-warner}
\begin{equation}
W= c_1 {\rm Tr}(B_i A_i)^2 + c_2{\rm Tr}(A_i B_i)^2
\end{equation}

This family of superpotentials includes both the ${\cal N}=3$ theory
and the SO(4) invariant ${\cal N}=2$ theory ($c_1 = - c_2 = c$).
What is the form of the RG flow in this two-dimensional space?

Consider the chiral ring for each choice of $c_i$. The relations in
the chiral ring $\partial W=0$ eliminate a single linear combination
$W$ of the two operators ${\rm Tr}(B_i A_i)^2$ and ${\rm Tr}(A_i
B_i)^2$. Hence there is a single dimension $2$ chiral operator and a
corresponding  marginal direction for each value of the $c_i$. This
indicates the existence of an exactly marginal line in the space of
couplings which passes through the ${\cal N}=3$ theory and the
SO(4) invariant ${\cal N}=2$ theory, and presumably connects them.

An alternative argument is that the condition for conformal
invariance of the superpotential is  $2 \gamma_{A_i} + 2
\gamma_{B_i}=0$, a single equation in the two variables $c_i$.

As for the four-dimensional theory with the same superpotential
\cite{Corrado:2002wx,corrado-halmagyi,
benvenuti-hanany,halmagyi-pilch-romelsberger-warner},
 one may generalize this to superpotentials which
break completely the continuous flavor symmetry, but preserves
enough discrete symmetry that the number of constraints from
conformal symmetry is lower than the number of coefficients in the
superpotential.

Examining the spectrum of protected operators in \cite[Table
1]{nilsson-pope} we find another interesting fact: there is a single
dimension $3$ operator ($p=1$) which is $Sp(2)$ invariant, from the
$(0,2,0)$ representation of $SO(6)$. This operator could be added to
the ${\cal N}=0$ Lagrangian to break the flavor symmetry to $Sp(2)$,
or to the ${\cal N}=1$ Lagrangian to break SUSY but preserve the
$Sp(2)$ flavor symmetry. This operator will have a small nonzero
anomalous dimension in either theories, controlled by the small
parameter $\lambda_1 + \lambda_2$, and it is a good candidate to
parameterize a slow flow between the ${\cal N}=0$ and the ${\cal
N}=1$ fixed points.


\subsection{Other quivers} 
\label{sub:quivers}

The logic in section \ref{sec:romans} can be applied to other AdS$_4$/CFT$_3$
duals. There are many AdS$_4$ of the Freund--Rubin type, and one
would expect that all of them should correspond to Chern--Simons--matter theories, with quivers obtained in various
ways. The procedure to obtain these theories is not yet as well
understood as its counterpart for four--dimensional theories.
Nevertheless, there is now an infinite series of
${\cal N}=3$ examples \cite{jafferis-t,imamura-kimura}, which can be derived using the same duality procedure as in \cite{aharony-bergman-jafferis-maldacena}. Interestingly,
proposals have been made for ${\cal N}=2$ \cite{martelli-sparks-3dquivers,
hanany-zaffaroni-cs,hanany-vegh-zaffaroni,kim-lee-lee-park}, and
${\cal N}=1$ duals \cite{ooguri-park}.

In particular, we can repeat the reasoning using brane probes for
 the ${\cal N}=3$ examples (for which one
can write an explicit Lagrangian). In that case, the quivers are
``necklaces'' consisting of $N_{\rm nodes}$ nodes, the $i$--th
being connected to the $(i+1)$--th and $(i-1)$--th by
arrows going both ways (for details see \cite{jafferis-t,imamura-kimura}). The levels should satisfy
\begin{equation}\label{eq:sumzero}
    \sum_{i=1}^{N_{\rm nodes}} k_i = 0 \ .
\end{equation}
So far for the original duals, which are of the Freund--Rubin
type, and thus in particular do not have any Romans mass $F_0$.
If one wants now to introduce a $F_0\neq 0$, along similar lines
as in section \ref{sec:romans} one arrives at the identification
\begin{equation}\label{eq:F0ki}
    F_0 = \sum_{i=1}^{N_{\rm nodes}} k_i \ .
\end{equation}

One can then apply the logic in this section, to see whether these
theories exist. The easiest theory to find is again the ${\cal N}=3$
one. In field theory, there is no particular reason to impose
(\ref{eq:sumzero}); that equation is valid for the theories whose
gravity dual is known. (\ref{eq:F0ki}) predicts that the more
general theories without the relation (\ref{eq:sumzero}) will have a
dual with non--zero Romans mass. These solutions are not known yet,
just as the dual to the ${\cal N}=3$ theory discussed in section
\ref{sub:ft3} (as we will see in section \ref{sub:gr3}).

We can also try to generate superconformal points with different
symmetries, along the lines of sections \ref{sub:ft0}, \ref{sub:ft1}
and \ref{sub:ft2}. Indeed if the Chern Simons levels are alternating
$k_i = (-1)^i k$, the necklace quivers enjoy ${\cal N}=4$
supersymmetry 
\cite{hosomichi-lee-lee-lee-park,benna-klebanov-klose-smedback}, with an $SO(4)$ R-symmetry. We could
then modify the levels slightly and consider a space of ${\cal N}=0$
theories which preserve the $SO(4)$ flavor symmetry, of ${\cal N}=1$
theories which preserve $SO(3)$ flavor symmetry, or ${\cal N}=2$
theories with $SO(2)_R \times SO(2)$ symmetry. There will also be
other exactly conformal ${\cal N}=2$ theory with more general
superpotentials. It would be interesting to develop this point
further.

\subsection{Multitrace deformations} 
\label{sub:Multitrace}

In this section we would like to study the effect of multitrace
terms on the existence of the four deformations of the ABJM CFT.

One could be concerned with the possibility that the existence of
marginal protected multi-trace operators will destroy the
perturbative construction of fixed points for the beta functions.
\begin{equation}\label{eq:iuvant}
    \beta_O = (\lambda_1 + \lambda_2) \delta \beta_O + \delta c_O \gamma_O\
\end{equation}

If the operator is marginal in the ABJM theory, $\gamma_O$ will also
be small, proportional to $\lambda_1 + \lambda_2$. On the other
hand, the planar expansion of the gauge theory gives us a hand:
$\delta \beta_O$ will contain extra factors of $1/N^2$, and $\delta
c_O$ will still be small at a fixed point.

Such marginal operators definitely exist. ${\rm Tr} X^I \bar X_J
{\rm Tr} X^J \bar X_I $ and ${\rm Tr} X^I \bar X_J {\rm Tr} X^K \bar
X_T \omega_{IK} \omega^{JT}$ are protected dimension $2$ double
trace operators in the ABJM theory which preserve the $Sp(2)$ flavor
symmetry of the ${\cal N}=1$ theory. It is also possible to write
several double trace Yukawa couplings and double or triple trace
potentials for the ${\cal N}=0$ theory which are protected in the
ABJM theory.

${\rm Tr} A^i B_u {\rm Tr}A^j B_v \epsilon_{ij} \epsilon^{uv}$ is a
chiral $SO(4)$ invariant operator in in the ${\cal N}=2$ theory.
Although perturbatively the superpotential is not renormalized, one
may imagine that non-perturbative effects in $\lambda_-$ may
introduce it. Notice that this operator is protected in the 
${\cal N}=2$ Lagrangian as well, not just in the ABJM theory. This means
that $\delta c_O$ will enter the $\beta$ function quadratically,
instead of linearly as in (\ref{eq:iuvant}),
possibly leading to $\delta c_O$ of order $1/N$.

Finally, there is one relevant double trace operator in the 
${\cal N}=0$ theory, ${\rm Tr} X^I \bar X_J {\rm Tr} X^J \bar X_I $.


\section{Gravity duals} 
\label{sec:gravity}

The conformal field theories we have described in section \ref{sec:multi} should all have gravity duals. We were able to find
duals only for the ${\cal N}=0$ and ${\cal N}=1$ cases, which
ironically are the ones with less control from the field theory
side.

Our electric basis will be
made of the internal fluxes, that is, the ones with no indices
in the spacetime, just as in \cite{gmpt3}. These include a
zero--form $F_0$ (the Romans mass), a two--form $F_2$, a
four--form $F_4$, a volume form $F_6$.

\subsection{${\cal N}=0$} 
\label{sub:gr0}

In this subsection\footnote{Some of the computations in this
section were done, in a different context, in conversations with Mariana Gra\~ na. The idea of finding solutions by
switching on only singlets of the internal SU(3) structure,
as we will do below, is of course not new; in massive IIA,
see \cite{lust-marchesano-martucci-tsimpis} for non--supersymmetric
examples, \cite{behrndt-cvetic} for supersymmetric ones,
and even \cite{romans-massiveIIA} for early examples.}, we will look for ${\cal N}=0$ solutions with
SO(6) symmetry. This means that we can just take the metric
to be the usual Fubini--Study metric on $\cc\pp^3$.

We will also assume that the dilaton is constant,
as well as the warping (the
function of the internal coordinates multiplying the AdS metric).
Actually, when the latter is constant,
it can be just set to zero by rescaling.
The equations of motion for the dilaton and metric
are then (in the string frame)
\begin{equation}\label{eq:EoM}
    R_{10}=\frac12 H^2\ ,\qquad
    R_{MN}-\frac12 g_{MN} R_{10}= g^2 T_{MN}\
\end{equation}
with
\begin{equation}
    \label{eq:TMN}
    \begin{array}{rl}\vspace{.2cm}
        T_{MN}=&\sum_{k=0}^3
        \frac12 \left(\frac1{(2k-1)!}F_{M P_1\ldots P_{2k-1}}F_N{}^{P_1\ldots P_{2k-1}}-\frac1{2\, (2k)!}
g_{MN}F_{P_1\ldots P_{2k}}F^{P_1\ldots P_{2k}}\right)+ \\
        &\frac1{2g_s^2}\left(\frac12 H_{MPQ} H_N{}^{PQ}-\frac1{12}g_{MN}F_{PQR}F^{PQR}\right)
        \ .
    \end{array}
\end{equation}

For a geometry AdS$_4\times M_6$ with $M_6$ Einstein, the
whole content of Einstein's equations is
in their four--dimensional trace
\begin{equation}\label{eq:R4}
    R_4= -g_s^2\sum_k F_k^2 \ , \qquad
\end{equation}
and their six--dimensional trace
\begin{equation}\label{eq:R6}
    R_6=\frac 12 \left( 3 H^2 + g_s^2 \sum_k (k-3)F_k^2 \right) \ ,
\end{equation}
where, again, the $F$'s are internal. The equations of motion
are, then, these two and the one for the dilaton (the first in
(\ref{eq:EoM})).
In fact, (\ref{eq:R4}) just sets
the value for the four--dimensional cosmological constant.
 It is also convenient to combine the first in (\ref{eq:EoM}) with
the sum of (\ref{eq:R4}) and (\ref{eq:R6}), to obtain
\begin{equation}\label{eq:H2}
    2 H^2 = g_s^2\sum_k (5-k)F_k^2\ .
\end{equation}

We also have to consider the equations of motion for the internal
fluxes $H$ and $F_k$,
and their Bianchi identity. If we assume that no sources are present,
these read
\begin{equation}\label{eq:bianchiEoM}
    d F= H \wedge F \ , \qquad d *F= -H\wedge F\ , \qquad d H=0 \ , \qquad d*H = -g_s^2\sum_{k=0}^2(F_{2k}\wedge *F_{2k+2}).
\end{equation}

So far, we have only used that the internal space is Einstein.
We now start using the fact that it is also \ka. We will consider an Ansatz for the fluxes
\begin{equation}
    \label{eq:sing}
    F=f_0\ , \qquad  F_2= f_2 J_0\ , \qquad F_4= f_4 \frac{J_0^2}2\ , \qquad F_6= f_6 \frac{J_0^3}6 \ , \qquad H= 0\ ,
\end{equation}
where $J_0$ is the \ka\ form and
the $f_i$ are taken to be constant.
The motivation for this Ansatz is simplicity: the idea is to use
nothing but singlets of U(3) (the structure group of the manifold).
In the supersymmetric case, a similar idea was used in \cite{behrndt-cvetic}; in the non--supersymmetric case, in
\cite{lust-marchesano-martucci-tsimpis}.

Let us now see whether we can find any solutions with this Ansatz.
The equations of motion and Bianchi identities for the internal
fluxes are (\ref{eq:bianchiEoM}). The first three of them are
trivially satisfied, because $H=0$ and $J_0$ is closed.
To evaluate the fourth, we use
\begin{equation}
    F\cdot F {\rm vol}=
    \frac1{k!}F_{m_1 \ldots m_k}F^{m_1 \ldots m_k}
    {\rm vol}=
    (-)^k F\wedge *F
\end{equation}
to compute
\begin{equation}
    \label{eq:Jcontr}
    1\cdot 1 = \frac{J_0^3}{6!}\cdot\frac{J_0^3}{6!}=1 \ ,
\qquad   J_0 \cdot J_0 =
    \frac{J_0^2}2\cdot \frac{J_0^2}2= 3 \ .
\end{equation}
Then the fourth in (\ref{eq:bianchiEoM}) gives
\begin{equation}\label{eq:ff}
    f_0 f_2 + 2 f_2 f_4 + f_4 f_6 = 0 \ .
\end{equation}

We now look at the equations of motion of the dilaton
and metric. For the internal metric, we use the conventions of section \ref{sub:gr1}; namely, we take the metric in (\ref{eq:1metric}), restricted to the case $\sigma=2$. This gives $R_6 = \frac{48}{R^2}$.  Using (\ref{eq:sing}) and (\ref{eq:Jcontr}),
(\ref{eq:R4}), (\ref{eq:R6}) and (\ref{eq:H2}) read
respectively
\begin{align}
    &\Lambda = -\frac{12}{R^2} \ ;
    \label{eq:Lambda}\\
    &\frac {48}{R^2}= \frac32 g_s^2 (-f_0^2 - f_2^2 + f_4^2 +f_6^2)\ ;
    \label{eq:Rg} \\
    &5 f_0^2 + 9 f_2^2 + 3 f_4^2 - f_6^2=0 \ .\label{eq:hf}
\end{align}
(\ref{eq:Lambda}) comes simply from $R_4 = 4 \Lambda$ and $R_{10}=0$ (which is the equation of motion for the dilaton, the first in (\ref{eq:EoM}), for $H=0$). Equations (\ref{eq:Lambda}) and (\ref{eq:Rg}) determine $\Lambda$ and $g_s$ in terms of
the flux parameters $f_{2k}$. 
The latter are constrained, however,
by (\ref{eq:ff}) and (\ref{eq:hf}). One can solve this system by
writing\footnote{One can divide both
equations by $f_6$, then use (\ref{eq:ff}) to express
$f_4/f_6$ in terms of $f_2/f_6$ and $f_0/f_6$, then
plug this into (\ref{eq:hf}) and solve for $f_0/f_6$.}
\begin{equation}\label{eq:sol0}
    f_0^2 (f_2^2+ 5 (2 f_2 + f_6)^2)=
    (f_6^2 -9 f_2^2)(2 f_6 + f_2)^2\ ,
\end{equation}
in the sense that one can use this equation
to determine $f_0$ in terms of $f_2$ and $f_6$.

Notice that for the $\nn=6$ solution, $f_6^2 = 36/(g_s^2 R^2)$ and
$f_2^2 = 12/(g_s^2 R^2)$, whereas $f_0=f_4=0$.

At this point, as far supergravity goes, we have found a family
of solutions, parameterized by $f_2$, $f_6$, and by $R$,
which at the moment is a free parameter. There is also the possibility
of adding a closed $B$ field, which can be parameterized by
the number
\begin{equation}\label{eq:b0}
    b\equiv \int_{\cc\pp^1} B\ .
\end{equation}
We now want to show that this family of solutions survive flux quantization and stringy corrections.

\subsubsection{Flux quantization} 
\label{sub:fq0}

As we already remarked in section \ref{sec:romans}, the
fluxes that couple to the brane worldsheets are
defined as $\tilde F_k \equiv (e^{-B} F)_k$; for example,
$\tilde F_2 \equiv F_2 - B F_0 $. The quantization law
then reads
\begin{equation}\label{eq:nft}
    \zz \ni n_{2k} = \int_{\cc\pp^k}\tilde F_{2k}\ .
\end{equation}
If we use the Ansatz (\ref{eq:sing}), we get that
\begin{equation}\label{eq:fq0}
    f_k = \frac{n^b_k} {R^k v_k}
\end{equation}
where $v_k$ are some numerical coefficients (the volumes
of the $\cc\pp^{k/2}$ for $R=1$), the $n^b_k$ are defined
as
\begin{equation}\label{eq:nb}
    \left(\begin{array}{c}\vspace{.3cm}
        n^b_0 \\ n^b_2\vspace{.3cm} \\ n^b_4 \vspace{.3cm}\\ n^b_6
    \end{array}\right)
    \equiv
    \left(\begin{array}{cccc}\vspace{.3cm}
        1 & 0 & 0 & 0\\\vspace{.3cm}
        b & 1 & 0 & 0\\\vspace{.3cm}
        \frac12 b^2 & b & 1 & 0 \\
        \frac16 b^3 &  \frac12 b^2 & b & 1
    \end{array}\right)
    \left(\begin{array}{c}\vspace{.3cm}
        n_0 \\\vspace{.3cm} n_2 \\\vspace{.3cm} n_4 \\ n_6
    \end{array}\right)\ ,
\end{equation}
and $b$ was defined in (\ref{eq:b0}). Notice that in section
\ref{sec:romans}, we called $n_2=k$ and $n_6=N$.

Hence, for each choice $\{ n_k \}$ of the four flux integers,
flux quantization gives us the four equations (\ref{eq:fq0}).
As we remarked after (\ref{eq:sol0}), the family of solutions
we had before imposing flux quantization has four parameters:
$f_2$, $f_6$, $R$ and $b$. This sounds promising, but in general
such a system of real
equations might or might not have solutions, depending on the coefficients and hence on the $n_k$.
But we do know that the system
has a solution, namely the ${\cal N}=6$ solution \cite{nilsson-pope,
sorokin-tkach-volkov-11-10}. Actually, the solutions we care
about in this paper are the ones which are small perturbations
of this ${\cal N}=6$ solutions. Let us think of (\ref{eq:fq0})
as a map from $(f_2, f_6, R, b)$ to $(n_0, n_2, n_4, n_6)$. All
we need to do, then, is to make sure that the image of this map
extends in the $n_0$ direction around the ${\cal N}=6$ solution.
This can be done explicitly in perturbation theory. If one just
includes a small $n_0 \ll n_2, n_6$, without including $n_4$
for simplicity, one finds $\delta f_0 = n_0$,
$\delta f_4=-\frac15\delta f_0$, $\delta b= \frac{R^4}{n_2}\delta f_4$.
This shows that the ${\cal N}=6$ solution is not isolated in the
$n_0$ direction. We have also found solutions of this type numerically.

Stringy corrections are also under control.
The solutions to these equations will be analytic in the coefficients, if one lets them vary continuously. So, once again,
if one starts from the $\nn=6$ solution, for which $R$ can be
made large and $g_s$ small,
and one adds some small amount of $n_0$, $R$ will be still large,
and $g_s$ will still be small.

Finally, one should wonder whether this solution is stable, since we no longer have supersymmetry to protect it. Fortunately, the ${\cal N}=6$ solution is not only stable (as it should be because of supersymmetry): its mass spectrum has also a gap above the Breitenlohner--Freedman bound, as one can again see from 
\cite[Table 1]{nilsson-pope}. In other words, not only do all the fields in the spectrum satisfy the bound: none saturates it. For ${\cal N}=0$ solutions that are small perturbations of the ${\cal N}=6$ solution, which are the ones we are interested in, the Breitenlohner--Freedman will then still be satisfied, since no mass can suddenly fall below the bound. 

In conclusion, we have found a set of solutions of IIA with
isometry group SU(4)$=$SO(6), with no preserved supersymmetry, and with non--zero Romans mass $F_0$. Using the logic in section \ref{sec:romans}, we propose that these solutions should be dual to the
theories discussed in subsection \ref{sub:ft0}, with the map
\begin{equation}\label{eq:match}
    k_1 = n_2 + n_0 \ ,\qquad k_2 = -n_2 \ ,\qquad
    N_1 = n_6 + n_4 \ ,\qquad N_2 = n_6  \ .
\end{equation}
at least when
$F_0$ is small with respect to the other flux quanta.\footnote{The reader might be unhappy about the seeming asymmetry of this formula.
Notice, however, that a shift of $B$ by the closed form with $2 \pi$
period will send $\tilde F_6\to \tilde F_6 + \tilde F_4$, which
(via (\ref{eq:nft}))
will exchange the role of $N_1$ and $N_2$ in (\ref{eq:match}).
At the same time, this will also exchange the role of $k_1$ and $k_2$.
Naively this would also generate contributions of order $k$
to $N_1$ and $N_2$; presumably the resolution to the puzzle comes
from the Riemann$^2$ 
terms in the brane action, which indeed should give
rise to corrections of order $(l_s/R)^4\sim k/N$. It would
be interesting to check this in detail.}


\subsubsection{Probe branes} 
\label{sub:probe}

In this section, we perform a probe brane analysis on the ${\cal N}=0$ vacua we just found. 

Consider a D2 extended along the three non--radial directions of spacetime. We want to ask whether such a brane is stable or unstable to expanding in the radial direction. This will tell us whether the potential of the dual theory is positive definite or not.   

The potential for such a D2 is 
\begin{equation}\label{eq:Vprobe}
	V= -\mu_2 \left[ \int d^3 \sigma \sqrt{-|g|} + \int C_3 \right]\ .
\end{equation}
Here, $\mu_2$ is the D2 tension. $|g|$ is the determinant of the (pull--back of) the spacetime metric; the relevant part is the standard AdS$_4$ metric $\frac{R^2_{AdS}}{r^2}( dr^2 + (dx^0)^2+(dx^1)^2+(dx^2)^2)$.  $C_3$ is the three--form potential. When we described the ${\cal N}=0$ vacua earlier in this section, recall that we chose the internal fluxes as our electric basis. Among which we had $F_6=f_6 {\rm vol}_6$. So far we have not used any of the ``magnetic'' fluxes $\tilde F_i$, namely the ones which have also legs along the spacetime. These can be obtained from the ``electric'' internal fluxes by duality; in particular, we have $\tilde F_4= f_6 {\rm vol}_{\rm AdS}$. Taking a potential $C_3$ with no leg in the radial direction, we have $dC_3 = dr \wedge \del_r C= \tilde F_4$, which gives us 
\begin{equation}
	C_3= -\frac{R^4_{\rm AdS}}{3 r^3} f_6 dx^0 \wedge dx^1 \wedge dx^2\ .
\end{equation}
Summing up, we have
\begin{equation}\label{eq:Vexp}
	V = -\frac{\mu_2}{r^3}\left( \frac{R^3_{\rm AdS}}{g_s} - \frac{f_6}3 R^4_{\rm AdS}\right)= 
	-\mu_2 \frac{R^3_{\rm AdS}}{r^3} \left( \frac1{g_s}- \frac R6 f_6\right)
\end{equation}
where we have used $R_{\rm AdS}= \sqrt{-\frac 3 \Lambda}$, and (\ref{eq:Lambda}). 

Notice that, as remarked earlier in this section, for the ${\cal N}=6$ solution we have $f_6=\frac 6{g_s R}$; hence, in that case, the potential in (\ref{eq:Vexp}) is identically zero. This is how it should be: the potential for the ${\cal N}=6$ CFT has flat directions that correspond to the D2 moving along $r$.
In the general case, one can just use (\ref{eq:Rg}), (\ref{eq:sol0}) and (\ref{eq:ff}) to express $V$ in (\ref{eq:Vexp}) as a (complicated) function of $R$, $f_2$ and $f_6$. Actually $R$ is just a multiplicative overall factor; one can also factor out $f_2$ and study the remaining function of $\frac{f_6}{f_2}$. Upon noticing from (\ref{eq:sol0}) that $\left| \frac{f_6}{f_2} \right| \ge 3$, one finds  (if $f_6$ and $f_2$ have equal sign) that $V \le 0$: the electric repulsion term wins over the gravitational attractive term. These ${\cal N}=0$ vacua are then non--perturbatively unstable towards nucleation of D2 branes. The dual field theories will then have an unstable potential. 



\subsection{${\cal N}=1$} 
\label{sub:gr1}

In this case, field theory instructs us to look for a family
of solutions on $\cc\pp^3$
with ${\cal N}=1$ supersymmetry, SO(5) symmetry,
and non--zero Romans mass.

In fact, such a family already exists \cite{t-cp3}. The metric
is no longer the Fubini--Study metric. Topologically, $\cc\pp^3$
is an $S^2$ fibration over $S^4$. The metric can
 be written as
\begin{equation}\label{eq:1metric}
    ds^2_6 = R^2\Big(\frac18
    (d x^i + \epsilon^{ijk}A^j x^k)^2 +
    \frac1{2 \sigma} ds^2_{S^4}\Big)
\end{equation}
where $x^i$ are such that $\sum_{i=1}^3 (x^i)^2=1$, $A^i$
are the components of an SU(2) connection on $S^4$
(with $p_1 =1 $), and $ds^2_{S^4}$ is the round metric on $S^4$
(with radius one).
$R$ is an overall radius.
For $\sigma=2$, (\ref{eq:1metric}) is the usual Fubini--Study
metric, whose isometry group is SO(6); in this case, the metric
has a coset structure ${\rm SU}(4)/{\rm U}(3)$.
For $\sigma\neq 2$, the
isometry group is simply the SO(5) of the base $S^4$. Actually,
the metric still comes from a coset: ${\rm Sp}(2)/{\rm Sp}(1)\times
{\rm U}(1)$. This latter fact was emphasized in \cite{koerber-lust-tsimpis}, who redid the computations in \cite{t-cp3}
using this coset structure.

At the level of supergravity, the ${\cal N}=1$ solutions found in \cite{t-cp3} are a family with four parameters: the two parameters
$R$ and $\sigma$ in the metric (\ref{eq:1metric}), the string
coupling $g_s$, and a parameter $b$ similar to the one defined in
(\ref{eq:b0}). The difference in this case is that supersymmetry
requires the NS curvature $H$ be non--zero (see \cite[Eq.~(2.2)]{t-cp3}). One can solve that
constraint by writing
\begin{equation}\label{eq:BJ}
    B=\frac m{\tilde m} J+ B_0\
\end{equation}
where $m$ and $\tilde m$ are two functions of $\sigma$, and
$B_0$ is a closed two--form. One can then define
\begin{equation}\label{eq:b1}
    b\equiv \int_{\cc\pp^1} B_0 \ .
\end{equation}

Once again, we have a family of solutions with four parameters.
The flux quantization conditions read
\begin{equation}\label{eq:fluxq}
    \left(\begin{array}{c}\vspace{.3cm}
        \frac 5{2 r g_s}m_0\\ \vspace{.3cm}
        4 \pi \frac r{g_s}\frac {(\sigma-1)}
        { (\sigma+2)}\\\vspace{.3cm}
        -\frac43 \pi^2 \frac{r^3}{g_s} m_0
        \frac{(\sigma-1)(1+ 2\sigma)}{\sigma^2 (\sigma+2)^2 }\\
        -\frac4{15} \pi^3 \frac{r^5}{g_s}
        \frac{(1+ 2 \sigma)(\sigma^2 -12 \sigma -4 )}
        {\sigma^2 (\sigma+2)^2}
    \end{array}\right)
    =
        \left(\begin{array}{c}\vspace{.3cm}
            n^b_0 \\ n^b_2\vspace{.3cm} \\ n^b_4 \vspace{.3cm}\\ n^b_6
        \end{array}\right)
\end{equation}
where $n^b_k$ are defined as in (\ref{eq:nb}), and
\begin{equation}
    m_0 \equiv\sqrt{(\sigma-2/5)(2-\sigma)}\ .
\end{equation}
The vector on the left of (\ref{eq:fluxq}) is nothing but
$(F_0,\int_{\cc\pp^1}\tilde F_2,\int_{\cc\pp^2}\tilde F_4, \int_{\cc\pp^3}\tilde F_6)$, where $\tilde F_k\equiv
(e^{-\frac m{\tilde m}J}F)_k$, just like in section \ref{sub:gr0}.


Now we have a system given by the four equations (\ref{eq:fluxq})
for the four parameters $R$, $\sigma$, $g_s$ and $b$. To show
that the system has solutions close to the ${\cal N}=6$
solution\footnote{Several solutions to this system were found
in \cite{t-cp3}; in that paper, however, there was no reason
to look for them particularly close to the ${\cal  N}=6$
solution.}, we have proceeded using perturbation theory,
introducing a small $n_0 \ll n_2, n_6$, along
the lines explained in section \ref{sub:gr0} for the
${\cal N}=0$ solutions. Once again, we have also found
numerical examples.

It is very natural to say that these solutions should be dual
to the theories discussed in \ref{sub:ft1}, with the same matching
of discrete parameters as in (\ref{eq:match}).


\subsection{${\cal N}=2$} 
\label{sub:gr2}

In this case, we cannot offer the gravity dual to the field
theories discussed in \ref{sub:ft2}. The reason is essentially
that the amount of symmetry is smaller: in both the ${\cal N}=0$
and ${\cal N}=1$ cases, the metric was homogeneous: the orbit of
the isometry group (respectively, SO(6)$=$SU(4) and SO(5)$=$Sp(2))
was the whole space. In the ${\cal N}=2$ case, the isometry group
is SO(4), and its orbits have codimension 1. This means that, this
time, one really has to solve differential equations in one variable.

Another difficulty is that the ${\cal N}=2$ solution we are
looking for is not going to be unlike the supersymmetric solutions
known so far, in a sense we now specify.
Supersymmetric solutions can be broadly divided in two classes.
In type IIA, there are two supersymmetry parameters;
for an ${\cal N}=1$ vacuum solution, they can be decomposed as
\begin{equation}\label{eq:N1eps}
    \begin{array}{l}\vspace{.3cm}
    \epsilon^1 = \zeta_+ \otimes \eta^1_+ + \zeta_- \otimes \eta^1_-
    \ ,\\
    \epsilon^2 = \zeta_- \otimes \eta^2_+ + \zeta_+ \otimes \eta^2_-
    \ ,
    \end{array}
\end{equation}
where $\zeta_+$ is a chiral four--dimensional spinor,
$\eta^{1,2}_+$ are two chiral six--dimensional spinors,
and $\zeta_-= (\zeta_+)^*$, $\eta^{1,2}_-= (\eta^{1,2}_+)^*$
to make sure $\epsilon^{1,2}$ are Majorana. If
$\eta^1$ and $\eta^2$ are proportional, the solution is said to be ``SU(3) structure''; if they are not, it is called \stt\ structure. These strange--sounding names
come from the general classification of supersymmetric vacua using
generalized complex geometry \cite{gmpt2,gmpt3}).
The ${\cal N}=1$ solutions
of \cite{t-cp3}, that we just reviewed in section \ref{sub:gr1},
are SU(3) structure; as are,
to the best of our knowledge, all known supersymmetric AdS$_4$
solutions.\footnote{It should be easy, however, to
generate new supersymmetric solutions beyond this class by acting
with a solution--generating technique, as in \cite{lunin-maldacena}.}

For ${\cal N}=2$ solutions\footnote{Here we are talking about
solutions with non--vanishing RR fields. Without this assumption,
the supersymmetry equations do not mix the $\epsilon^1$ and
$\epsilon^2$, and it is
possible to achieve ${\cal N}=2$ by having two different $\zeta$'s
in the two rows of (\ref{eq:N1eps}).}, one needs two four--dimensional
spinors, and as a consequence we need another pair of internal
six--dimensional spinors too:
\begin{equation}\label{eq:N2eps}
    \begin{array}{l}\vspace{.3cm}
    \epsilon^1 = \sum_{a=1}^2(\zeta^a_+ \otimes \eta^{1,\,a}_+ +
                + \zeta^a_- \otimes \eta^{1,\,a}_- ) \equiv
                \sum_{a=1}^2 \epsilon^{1,\,a}
    \ ,\\
    \epsilon^2 = \sum_{a=1}^2(\zeta^a_+ \otimes \eta^{2,\,a}_+ +
                + \zeta^a_- \otimes \eta^{2,\,a}_- ) \equiv
                \sum_{a=1}^2 \epsilon^{1,\,a}
                \ .
    \end{array}
\end{equation}
In this case, the pair $\eta^{1,\,a=1}, \eta^{2,\,a=1}$ and the pair
$\eta^{1,\,a=2}, \eta^{2,\,a=2}$ have each to solve the same equations as in the ${\cal N}=1$ case; but with the same metric and fluxes. Also, the R--symmetry generator $R$ sends the first solution into the second.
That implies that $\epsilon^{i,\,a=1}=R^i{}_j \epsilon^{j,\,a=2}$
for some matrix $R^i{}_j$. Hence, the pair for $a=1$ is
SU(3), so will be the pair for $a=2$. For this reason, it still
makes sense to divide solutions into SU(3) and \stt, like we did
in the ${\cal N}=1$ case.

SU(3)--structure solutions are characterized
by a set of equations first found in \cite{lust-tsimpis}; in the
present notation, they can be found in \cite[Eq.~(2.2,2.6)]{t-cp3}.
To have an ${\cal N}=2$ solution with non--zero $F_0$, one needs
to find two solutions $(J_1,\Omega_1)$ and $(J_2,\Omega_2)$ to those
equations; the index is the $a=1,2$ that we saw in (\ref{eq:N2eps}).
As we mentioned above, these two pairs should solve the equations
with the same fluxes $H$, $F_{2k}$ and the same
metric $g$. The latter is determined as
\begin{equation}\label{eq:gji}
    g=J_1 I_1= J_2 I_2
\end{equation}
 where $I_a$ are two almost
complex structures such that $\Omega_a$ are their $(3,0)$
forms. Now, for $F_0\neq 0$, one finds from \cite[Eq.~(2.2)]{t-cp3}
that\footnote{This equation is also related to (\ref{eq:BJ}).}
\begin{equation}
    H = \frac25 g_s F_0 {\rm Re} \Omega\ .
\end{equation}
This equation should be true for both $\Omega_a$. Hence one gets that
${\rm Re} \Omega_1 = {\rm Re} \Omega_2$. But both $\Omega_a$
should be decomposable (namely, locally the wedge of three
one--forms), because they should be $(3,0)$ forms with respect to
same almost complex structures $I_a$. This implies that
each ${\rm Re} \Omega_a$ should determine the whole of
$\Omega_a$ (see for example \cite{hitchin-67}); hence
one gets $\Omega_1= \Omega_2$.
But a decomposable non--degenerate form
$\Omega$
actually determines an almost complex structure $I$ under which it should be a $(3,0)$--form. This means that also $I_1=I_2$. Finally, because of
(\ref{eq:gji}), $J_1=J_2$. So the two solutions to the
supersymmetry equations are actually the same.

In summary, we have shown in full generality that a
supergravity solution with extended supersymmetry
and $F_0 \neq 0$ cannot be SU(3)--structure.

This result is a complication for us: the equations for the \stt\ case
are far more complicated than the ones for the SU(3) case. Although both SU(3) and \stt\ are particular cases of \cite[Eq.~(7.1), (7.2)]{gmpt3}, written in the language of
generalized complex geometry, the \stt\ case
needs to be massaged significantly before
they can be applied to any concrete situation. Although work is
in progress on this, we do not have as yet a solution
to offer.

The fact that the solution has to be \stt\ structure can also
be seen purely from field theory, as we now proceed to show.
From the general rules of the AdS/CFT correspondence, we expect
to be able to compare domain wall D--branes in AdS$_4$, extended
along the subspace defined by $r=$const in Poincar\'e coordinates,
with vacua of the CFT. Also, BPS domain walls in AdS$_4$
 should correspond to supersymmetric vacua in the CFT.

For the particular CFT described in
section \ref{sub:ft2}, the gauge group is U$(N)\times$U$(N)$,
and $N$ should be large for the $\alpha'$ corrections to be under
control on the gravity side; but let us imagine we are only giving expectation value to a 1$\times$1 block. In this abelian case, the moduli space of supersymmetric vacua is given by
\begin{equation}\label{eq:susyvacuaN2}
    \sum_{i=1}^2 (|A_i|^2 - |B_i|^2)= 0 \ ,
\end{equation}
since the superpotential in (\ref{eq:cft2}) vanishes in this abelian
case.

One also has to take care of possible actions on the moduli space
by the gauge group.  
The moduli space of the original U(1)$\times$U(1) 
ABJM theory is a $\cc^4 /Z_k$ parameterized by
the $A_i, \bar B_i$. The $A_i, \bar B_i$ fields all have the same
charge under the linear combination  ${\cal A}_1 - {\cal A}_2$ of
the two U(1) gauge fields, so the naive Higgsing would give as a
moduli space the cone over $\cc \pp^3$. The particular choice of
opposite Chern--Simons levels replaces the naive U(1) gauging by
the discrete $\zz_k$ quotient, to give $\cc^4/\zz_k$. 

If the sum of the levels is not zero, the U(1) gauging is no
longer discretized, and it happens in
full. The moduli space of a ${\cal N}=2$ theory should be \ka,
and indeed (\ref{eq:susyvacuaN2}) is the correct equation for a 
\ka\ quotient of $\cc^4$.

Let us now look at the gravity side.
The BPS conditions for domain walls can be written conveniently
\cite{martucci} in terms of the pure spinors $\Phi_\pm$ of
generalized complex geometry, that also appear in the supersymmetry
equations for the background \cite{gmpt3}. We expect the abelian
vacua in the CFT should correspond to a D2 domain wall, situated at
$r=$const and pointlike in the internal $\cc\pp^3$.\footnote{Domain
walls wrapping higher--dimensional subspaces of $\cc\pp^3$ will
presumably come from Myers--like effects in the CFT.} For such a
brane, in the conventions of \cite[Eq.~(7.1), (7.2)]{gmpt3} the only
condition is
\begin{equation}\label{eq:bpsD2}
    {\rm Im} (e^{i\theta} \Phi^a_{+(0)})_| =0
\end{equation}
where $e^{i\theta}$ is the phase of $\mu$, $\Phi_{+(0)}$ is
the zero--form part of the $\Phi_+$ pure spinor, $a$ is the
index introduced in (\ref{eq:N2eps}),
and ${}_|$ denotes
restriction to the point in $\cc\pp^3$ where the D2 is located.
Now, the crucial point is that, in the SU(3) case, it was shown
in \cite[Sec.~7]{gmpt3} that the left hand side of (\ref{eq:bpsD2})
is constant. This means that (\ref{eq:bpsD2}) either has solution
everywhere on the $\cc\pp^3$, or nowhere. This does not match
with what we found on the CFT side, namely the single equation (\ref{eq:susyvacuaN2}).

In the \stt\ case, there is no reason for the left hand side of
(\ref{eq:bpsD2}) to be constant. The equation (\ref{eq:bpsD2})
should then just match (\ref{eq:susyvacuaN2}). Indeed, it is easy to
argue from the isometries of the problem that
$\Phi^1_{+(0)}=\Phi^2_{+(0)}$ is constant on the orbits of the
$SO(4)$ flavor symmetry, which are given by
\begin{equation}\label{eq:susyvacuaN2bis}
    \sum_{i=1}^2 (|A_i|^2 - |B_i|^2)= t \ .
\end{equation}
The field theory predicts that the \ref{eq:bpsD2} will be satisfied
at $t=0$.

It remains to interpret the U(1) gauging from gravity. First of all, 
in the ABJM case, we saw that the moduli space of the field theory is $\cc^4/\zz_k$. From the point of view of the D2 domain walls, 
the combination of the radial
motion in $AdS_4$ and the motion in $\cc\pp^3$ gives a
cone over $\cc \pp^3$; an extra U(1)
circle arises from dualization of the wordvolume gauge field, 
and one reproduces this way the $\cc^4/\zz_k$ moduli space.

In the case with unequal levels, we saw instead that an U(1)
gauging happens, which is complexified by the condition (\ref{eq:susyvacuaN2}). 
From the point of view of the D2 domain wall, the $F_0$ background
induces a Chern--Simons coupling on the worldvolume, and the
worldvolume gauge field becomes massive, together with a transverse
scalar, which is its ${\cal N}=2$ supersymmetric partner.


\subsection{${\cal N}=3$} 
\label{sub:gr3}

This case is very similar to the ${\cal N}=2$ case. As for that
case, we cannot offer a gravity solution. And again to our partial
excuse, we can
apply the gravity argument we saw in section \ref{sub:gr2},
and conclude that the solution must be \stt.

The comparison of D2 BPS domain walls with vacua of the CFT
also goes along similar lines.
The BPS equation reads again (\ref{eq:bpsD2}), but with
$a$ going from 1 to 3. This should correspond to the three equations
for supersymmetric vacua for (\ref{eq:cft3}), that read
\begin{equation}
    \sum_{i=1}^2 (|A_i|^2 - |B_i|^2)= 0 \ ,\qquad
    \sum_{i=1}^2 A_i B_i =0 \ ,
\end{equation}
or, more symmetrically, $\sum_{i=1}^2 X_i^\dagger\sigma_\alpha X_i$,
with $X_i={A_i \choose B_i}$, and $\sigma_\alpha$ Pauli matrices.

These are the conditions for the U(1) hyper--\ka\ quotient of
$\cc^4$. In this case $\Phi^a_{+(0)}$ should be constant on the orbits
of the SO(3) flavor symmetry, which are given by
\begin{equation}
    \sum_{i=1}^2 (|A_i|^2 - |B_i|^2)= t_R \ ,\qquad
    \sum_{i=1}^2 A_i B_i =t_C \ .
\end{equation}
The field theory predicts that the \ref{eq:bpsD2} will be satisfied
at $t_R=t_C=0$.


\bigskip

{\bf Acknowledgments.} We would like to thank G.~Moore, J.~Maldacena and A.~Zaffaroni for interesting discussions. 
D.~G.~is supported in part by the DOE
grant DE-FG02-90ER40542 and in part by the Roger Dashen membership
in the Institute for Advanced Study. 
A.~T.~is supported in part by DOE grant DE-FG02-91ER4064.


\providecommand{\href}[2]{#2}

\end{document}